\documentclass[twocolumn,showpacs,preprintnumbers,amsmath,amssymb]{revtex4}

\usepackage{graphicx}
\usepackage{dcolumn}
\usepackage{bm}
\def\jpsi{{J/\psi}}
\def\hats{{\hat{s}}}
\def\hatscc{{\hat{s}_{c\overline{c}}}}
\def\OS{\mathrm{OS}}
\def\li{{\mathrm{Li_2}}}

\def\be{\begin{equation}}
\def\ee{\end{equation}}
\def\bea{\begin{eqnarray}}
\def\eea{\end{eqnarray}}
\def\NO{\nonumber}
\def\gev{\mathrm{~GeV}}
\def\kev{\mathrm{~KeV}}
\def\fb{\mathrm{~fb}}
\def\pb{\mathrm{~pb}}

\def\dfrac{\displaystyle\frac}
\def\co{{\cal O}}
\def\a{\alpha}
\def\b{\beta}

\def\d{\delta}
\def\e{\epsilon}
\def\g{\gamma}
\def\s{\sigma}

\begin{document}


\title{Next-to-Leading-Order QCD Corrections to $e^+e^-\rightarrow \jpsi c \overline{c}$ at the B Factories}%

\author{Bin Gong and Jian-Xiong Wang}%
\affiliation{
Institute of High Energy Physics, CAS, P.O. Box 918(4), Beijing, 100049, China. \\
Theoretical Physics Center for Science Facilities, CAS, Beijing, 100049, China.
}%
\date{\today}

\begin{abstract}{\label{abstract}}
We calculate the next-to-leading-order (NLO) QCD correction to $e^+e^-\rightarrow \jpsi c \bar{c}$ at the {\it B} factories, and present theoretical predictions on the momentum and production angular distribution for $J/\psi$ production, and momentum distribution for $J/\psi$ polarization at NLO for the first time.  By applying Brodsky, Lepage and Mackenzie scale setting for the renormalization scale, it is found that the QCD perturbative expansion is significantly improved with the unique scale choice $\mu^*=1.65$GeV. Together with the $\psi^\prime$ feed-down contribution, the total cross section and momentum distribution can account for the recent experimental measurement by the Belle collaboration. The total cross section and momentum distribution are also found to be consistent with the experimental measurement in the previous study on $e^+e^-\rightarrow\jpsi gg$.  However, the production angular distribution of $\jpsi$ production for either the $\jpsi c\bar{c}$ or the $\jpsi gg$ channel has a quite different shape in contrast with the new experimental data, although it fits with the experimental data when the two channels are added together. This situation is difficult to explain. 
To clarify the puzzle of $J/\psi$ polarization, further experimental measurements are strongly expected to testify our predictions on the momentum distribution for $J/\psi$ polarization. Our total cross section agrees with that given in the previous study of Zhang and Chao by using their renormalization scheme and input parameters.
\end{abstract}

\pacs{12.38.Bx, 13.66.Bc, 14.40.Gx}
\maketitle
\section{Introduction}
For heavy-quarkonium production and decay, a naive perturbative QCD and nonrelativistic factorization treatment is applied straightforwardly. It is called color-singlet mechanism (CSM)~\cite{j.h.kuhn:79}. To describe the huge discrepancy of the high-$p_t$ $J/\psi$ production between the theoretical calculation based on CSM and the experimental measurement at Tevatron, a color-octet mechanism~\cite{Braaten:1994vv} was proposed based on the nonrelativistic QCD (NRQCD)~\cite{Bodwin:1994jh}. In the application, $\jpsi$ related productions or decays are very good places for two reasons, theoretically charm quark is thought to be heavy enough and charmonium can be treated within the NRQCD framework, experimentally there is a very clear signal to detect $\jpsi$. 
Now the integrated luminosity is more than $850\fb^{-1}$ at the Belle
detector at the KEKB and it is about 20 times larger than the
integrated luminosity $32.4\fb^{-1}$, based on which many
$\jpsi$ production processes was observed~\cite{Abe:2001za,Abe:2002rb,Aubert:2005tj}.
Therefore it supplies a very important chance to perform
systematical study on $\jpsi$ production both theoretically and
experimentally. A recent review on the situation can be found in Ref.~\cite{Brambilla:2004wf}.

The measurements for exclusive $\jpsi$ productions $e^+e^-
\rightarrow \jpsi  \eta_c$, $\jpsi  \jpsi$, $\jpsi
\chi_{cJ}$ at the B factories have shown that
there are large discrepancies between the leading-order (LO)
theoretical
predictions~\cite{Braaten:2002fi,Liu:2002wq,Hagiwara:2003cw,Bodwin:2002fk}
in NRQCD and the experimental
measurements~\cite{Abe:2002rb,Aubert:2005tj,Abe:2004ww}. It seems
that such discrepancies can be resolved by introducing higher order
corrections~\cite{Braaten:2002fi,Zhang:2005ch,Gong:2007db,Gong:2008ce,Bodwin:2002kk,Bodwin:2006yd,Bodwin:2006ke,He:2007te}:
next-to-leading-order (NLO) QCD corrections and relativistic
corrections.

The cross section for inclusive $\jpsi$ production in $e^+e^-$ annihilation was measured by BABAR~\cite{Aubert:2001pd,Aubert:2005tj} as $2.54\pm0.21\pm0.21\pb$ and Belle~\cite{Abe:2001za, Abe:2002rb} as $1.45\pm0.10\pm0.13\pb$. Many theoretical studies~\cite{Keung:1980ev,Driesen:1993us,Yuan:1996ep, Cho:1996cg,Schuler:1998az,Baek:1998yf,Liu:2002wq,Hagiwara:2007bq, Braaten:1995ez,Wang:2003fw} have been performed on this production at LO in NRQCD and the results for inclusive $J/\psi$ production cover the range $0.6\sim1.7\pb$ depending on parameter choices. 
A further analysis by Belle~\cite{Abe:2002rb} gives \be \sigma[e^+e^-\rightarrow \jpsi + c\bar{c}+X]=0.87^{+0.21}_{-0.19}\pm 0.17\pb, \ee 
and hints $\sigma[e^+e^-\rightarrow \jpsi+X({\mathrm{non}~c\bar{c}})]\sim0.6\pb$.
It is suggested in Ref.~\cite{Ioffe:2003gj} that different $\jpsi$ production mechanisms
can be tested by measuring $\jpsi$ polarization.
And a study of $J/psi$ polarization in $B\rightarrow \jpsi+ X$ with the BELLE detector presented in 
Ref.~\cite{Ichizawa:2000yh}.

For the non $c\bar{c}$ part, the contributions from the color-singlet channel $e^+e^-\rightarrow\jpsi gg$ and color-octet channel $e^+e^-\rightarrow\jpsi g$ are about $0.2$ and $0.27\pb$, respectively, at the LO in NRQCD~\cite{Wang:2003fw}. 
However, the signal of the color-octet was not found in the
experiment~\cite{Aubert:2001pd,Abe:2001za}. Therefore, the
experimental measurement by Belle is about 3 times larger than the
theoretical prediction from the color-singlet at LO, and can be much
more than 3 times by BABAR. The NLO QCD corrections to $e^+e^-\rightarrow\jpsi gg$ has been studied by two individual groups recently~\cite{Ma:2008gq,Gong:2009kp}, which boost the cross section to $0.373 - 0.496$ pb depending on parameter choices. Meanwhile, a new measurement with higher statistics reported by Belle~\cite{pakhlov:2009nj} gives
\be 
\sigma[e^+e^-\rightarrow \jpsi +X(\mathrm{non}~c\bar{c})]=0.43\pm0.09\pm0.09\pb,
\ee
which fits well with color-singlet predictions at NLO.

For the $c\bar{c}$ part, the experimental data by Belle~\cite{Abe:2002rb}, $0.87^{+0.21}_{-0.19}\pm 0.17\pb$, is about 5 times larger than the LO NRQCD prediction~\cite{Liu:2002wq}. However, this large discrepancy was partially resolved by considering both NLO correction and feed-down from higher excited states \cite{Zhang:2006ay}. It is also pointed out in Ref.~\cite{Nayak:2007mb} that the color transfer in associated heavy-quarkonium production may give important contribution to this process. The recent experimental measurement~\cite{pakhlov:2009nj} gives
\be 
\sigma[e^+e^-\rightarrow \jpsi +c\bar{c}]=0.74\pm0.08^{+0.09}_{-0.08}\pb, 
\ee
which is even closer to the theoretical predictions. Since the calculation of the NLO QCD correction to this process is quite complicated and plays a very important role to explain the experimental data, it is desirable to have an independent calculation. A more important point is that there are already the momentum and production angular distribution for $J/\psi$ production obtained in the new measurement to be compared with theoretical predictions. Furthermore, the transverse momentum distributions of $J/\psi$ polarization measured by CDF Collaboration~\cite{Abulencia:2007us} are still challenging our understanding of the heavy-quarkonium production mechanism even with the recent significant theoretical progresses~\cite{Campbell:2007ws,Gong:2008hk} on the NLO QCD calculation. To understand the $J/\psi$ polarization puzzle, there are also other $\jpsi$ related production processes calculated~\cite{Li:2008ym}, and it is very helpful to study $J/\psi$ polarization in $e^+e^-\rightarrow \jpsi +c\bar{c}$ from both theoretical and experimental parts. Therefore, in this paper, we present detailed study on the NLO QCD correction to $e^+e^-\rightarrow \jpsi +c\bar{c}$ by using our Feynman Diagram Calculation (FDC) package~\cite{FDC}, and give theoretical predictions on the momentum and production angular distribution for production, and momentum distribution for $J/\psi$ polarization at NLO for the first time. Our total cross section is in agreement with the previous result in Ref.~\cite{Zhang:2006ay} when their renormalization scheme and input parameters are used.

This paper is organized as follows. In Sec. II, we give the LO cross section for the process. The calculation of NLO QCD corrections is described in Sec. III. In Sec. IV, numerical results are presented. Further discussion on the renormalization scale choice is performed in Sec. V. The summary and discussions are given in Sec. VI. In the Appendix A, a trick which will bring better convergence in numerical calculation for some results is introduced.

\section{LO Cross Section}
\begin{figure}
\center{
\includegraphics[scale=0.40]{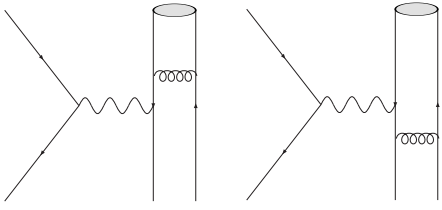}
\caption {\label{fig:LO}Typical Feynman diagrams at LO.}}
\end{figure}
The related Feynman diagrams which contribute to the LO amplitude of the process $e^+(p_1) + e^-(p_2) \rightarrow \jpsi(p_3) + c(p_4) + \bar{c}(p_5)$ are shown
in Fig.~\ref{fig:LO}, while the others can be obtained by reversing the arrows of quark lines. 
In the nonrelativistic limit, using the NRQCD factorization formalism, the differential cross section is obtained in $n=4-2\e$ dimensions as

\begin{widetext}
\bea
\dfrac{\mathrm{d}\s^{(0)}}{\mathrm{d}E_{\jpsi}}&=&
\dfrac{4\a^2\a_s^2e_c^2|R_s(0)|^2}{27m_c^6\hats^{5/2}}\biggl\{
\biggl[-\dfrac{128(2\hats+1)(\hats-1)^2\hats^2}{(\hatscc-1+\hats)^6}
+\dfrac{32(6\hats-1)(2\hats+1)(\hats-1)\hats}{(\hatscc-1+\hats)^5}
-\dfrac{8(8\hats-1)(2\hats+1)\hats}{(\hatscc-1+\hats)^4} \NO\\
&&
+\dfrac{4(8\hats^3 + 12\hats^2 + 3)}{\hats(\hatscc-1+\hats)^3}
-\dfrac{52\hats^4 - 30\hats^3 - 42\hats^2 - \hats + 9}{(\hats-1)\hats^2(\hatscc-1+\hats)^2}
+\dfrac{2(10\hats^5 - 18\hats^4 + 18\hats^3 - 5\hats^2 - 2\hats + 3)}{(\hats - 1)^2\hats^3(\hatscc-1+\hats)} \NO\\
&&
-\dfrac{2(4\hats + 3)}{\hats^3(\hatscc-1-\hats)}
+\dfrac{(2\hats + 3)(2\hats+1)}{\hats^2(\hatscc-1-\hats)^2}
-\dfrac{4(\hats + 2)}{(\hats -1)^2\hatscc}\biggr]\Delta_1\Delta_2\Delta_3
+\biggl[
-\dfrac{2(8\hats^3 + 2\hats^2 + 10\hats - 3)}{\hats(\hatscc -1+\hats)^2} \NO\\
&&
+\dfrac{6(6\hats^3 - 3\hats^2 + 1)}{\hats^2(\hatscc -1+\hats)}
+\dfrac{(2\hats^3 + 11\hats^2 - 6)}{\hats^2(\hatscc -1-\hats)}
-\dfrac{2(4\hats^2 - 3)}{\hats(\hatscc -1-\hats)^2}-\dfrac{2(2\hats+3)(2\hats+1)}{(\hatscc -1-\hats)^3}-16\biggr]\NO\\
&&
\times\ln\biggr(\dfrac{\hats+1-\hatscc+\Delta_1\Delta_3 /\Delta_2}{\hats+1-\hatscc-\Delta_1\Delta_3 /\Delta_2}\biggr) \biggr\} + \co(\e),
\label{eqn:E_LO}
\eea
\end{widetext}

where $s$ is the squared center-of-mass energy, $e_c$ and $m_c$ are the electric charge and mass of the charm quark, respectively. The dimensionless kinematic variables are defined as 
\bea
\hats&=&\dfrac{s}{4m_c^2}, \qquad ~~~ \hatscc=\dfrac{(p_4+p_5)^2}{4m_c^2}, \NO\\
\Delta_1&=&\sqrt{\hatscc-1}, \quad \Delta_2=\sqrt{\hatscc}, \NO\\
\Delta_3&=&\lambda^{1/2}(\hats,\hatscc,1)\equiv\sqrt{(\hats-\hatscc-1)^2-4\hatscc}.
\eea 
$R_s(0)$ is the radial wave function at the origin of $\jpsi$. The approximation $M_{\jpsi}=2m_c$ is taken. Our results at LO are consistent with those in Refs.~\cite{Hagiwara:2003cw,Zhang:2006ay}  
\section{NLO Cross Section}
At NLO in $\a_s$, there are virtual corrections from
loop diagrams. Dimensional regularization has been adopted for
isolating the ultraviolet (UV) and infrared (IR) singularities.
UV-divergences from self-energy and triangle diagrams are canceled
upon the renormalization of QCD. 
Here we adopt same renormalization scheme as Ref.~\cite{Gong:2007db}.
The renormalization constants $Z_m$, $Z_2$ and $Z_3$, which correspond to charm quark mass $m_c$, charm field $\psi_c$, and gluon field $A^a_\mu$ are defined in the on-mass-shell (OS) scheme while $Z_g$ for the QCD gauge coupling $\alpha_s$ is defined in the modified-minimal-subtraction($\overline{\mathrm{MS}}$) scheme:
\bea
\delta Z_m^{\mathrm{OS}}&=&-3C_F\dfrac{\alpha_s}{4\pi}\left[\dfrac{1}{\e_{UV}} -\gamma_E +\ln\dfrac{4\pi \mu^2}{m_c^2} +\frac{4}{3} \right] , \NO\\
\delta Z_2^{\OS}&=&-C_F\dfrac{\alpha_s}{4\pi}\biggl[\dfrac{1}{\e_{UV}} +\dfrac{2}{\e_{IR}} -3\gamma_E +3\ln\dfrac{4\pi \mu^2}{m_c^2} +4 \biggr] , \NO\\
\delta Z_3^{\OS}&=&\dfrac{\alpha_s}{4\pi}\biggl[(\beta'_0-2C_A)\left(\dfrac{1}{\e_{UV}} -\dfrac{1}{\e_{IR}}\right)
\NO\\&&
-\dfrac{4}{3}T_F\left(\dfrac{1}{\e_{UV}} -\gamma_E +\ln\dfrac{4\pi \mu^2}{m_c^2}\right) \biggr] , \NO\\
\delta Z_g^{\overline{\mathrm{MS}}}&=&-\dfrac{\beta_0}{2}\dfrac{\alpha_s}{4\pi}\left[\dfrac{1}{\e_{UV}} -\gamma_E +\ln(4\pi) \right] . 
\eea
where $\mu$ is the renormalization scale, $\g_E$ is Euler's constant, $\b_0=\frac{11}{3}C_A-\frac{4}{3}T_Fn_f$ is the one-loop coefficient of the QCD beta function and $n_f$ is the number of active quark flavors. There are three massless light quarks $u, d, s$, and one heavy quark $c$, so $n_f$=4. In $SU(3)_c$, color factors are given by $T_F=\frac{1}{2}, C_F=\frac{4}{3}, C_A=3$. And $\b'_0\equiv\b_0+(4/3)T_F=(11/3)C_A-(4/3)T_Fn_{lf}$ where $n_{lf}\equiv n_f-1=3$ is the number of light quarks flavors. 
Actually in the NLO total amplitude level, the terms proportion to $\delta {Z_3}^{\OS}$  cancel each other, thus the result is independent of renormalization scheme of the gluon field. 
\begin{figure*}
\center{
\includegraphics*[scale=0.7]{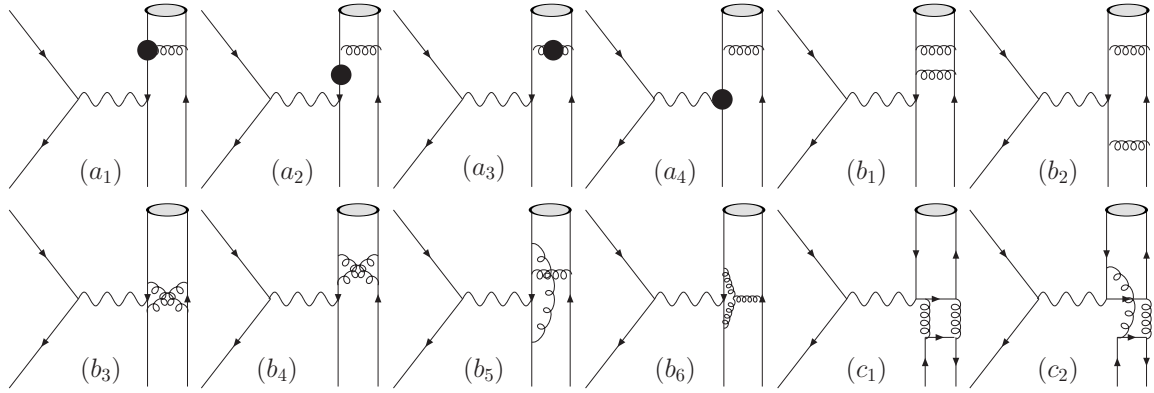}
\caption {\label{fig:NLO}Typical Feynman diagrams for virtual corrections. Groups $(a_1)-(a_4)$ are the counter-term diagrams, including corresponding loop diagrams. More diagrams can be obtained by reversing the arrows of the quark lines, and exchanging the places of the $J/\psi$ and open charm pairs in groups $(a)$ and $(b)$.}}
\end{figure*}

After having fixed our renormalization scheme and omitting diagrams that do not contribute, including counter-term diagrams, there are 80 NLO diagrams remaining, which are shown in Fig.~\ref{fig:NLO}.
Diagrams of groups $(b_1)$ and $(b_2)$ that have a virtual gluon line connected with the charm quark pair in $J/\psi$ lead to Coulomb singularity $\sim \pi^2/v$, which can be isolated by introducing a small relative velocity $v=|\vec{p}_{c}-\vec{p}_{\bar{c}}|$ and mapped into the $c\bar{c}$ wave function:
\bea
\s&=&|R_s(0)|^2\hat{\s}^{(0)}\left( 1 +\dfrac{\a_s}{\pi}C_F\dfrac{\pi^2}{v} + \dfrac{\a_s}{\pi}C +\co(\a_s^2)\right)  \NO\\
&\Rightarrow&|R^{ren}_s(0)|^2 \hat{\s}^{(0)} \left[1+ \dfrac{\a_s}{\pi}C +\co(\a_s^2)\right].
\eea

Although the Feynman diagrams are similar, the calculation of tensor and scalar
integrals is much more complicated than that in Ref.~\cite{Gong:2007db}. Again, the calculation was done automatically
with our FDC package\cite{FDC}.

\begin{figure}
\center{
\includegraphics[scale=0.5]{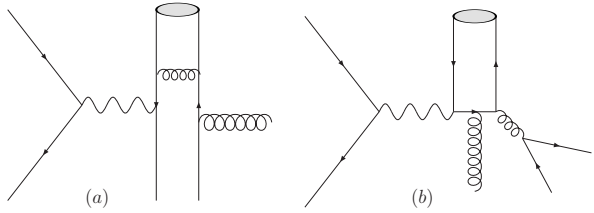}
\caption {\label{fig:real}Typical Feynman diagrams for real correction process $e^+e^-\rightarrow \jpsi c\overline{c}g$. $(a)$: Diagrams obtained by adding a gluon to LO diagrams. This type contains 24 diagrams; $(b)$: Diagrams with two quark lines. This type contains six diagrams. }}
\end{figure}
The real corrections arise from a real gluon emission process, $e^+e^- \rightarrow \jpsi c\overline{c}g$. There are two types of diagrams in this process, as shown in Fig.~\ref{fig:real}.
Usually phase space integration for real correction processes will generate IR singularities,
which is either soft or collinear and can be conveniently isolated
by slicing the phase space into different regions. We adopt the two-cutoff
phase space slicing method \cite{Harris:2001sx} to decompose the phase space into three parts by introducing two small cutoffs, $\d_s$ and $\d_c$. In this process, there is no collinear singularities and only the cutoff $\d_s$ is needed.
Then the real cross section is written as
\be
\s^R=\s^S+\s^{H\overline{C}},
\ee
where $\s^S$ from the soft regions contains soft singularities and is calculated analytically under soft approximation.
It is easy to find that the soft singularities for a gluon emitted from the
charm quark pair in the S-wave color singlet $\jpsi$ are canceled
by each other. We have
\bea
d\s^S&=&d\s^{(0)} \frac{\a_s}{2\pi} \frac{\Gamma(1-\e)}{\Gamma(1-2\e)}
\left(\frac{4 \pi \mu^2}{s}\right)^{\e}
\left(\dfrac{A_1}{\e} + A_0\right).
\eea
Two different frames are used to realize the division of the soft and hard noncollinear parts. 
The first one is in the center-of-mass (CM) frame of initial state particles $e^+e^-$, where $A_1$ and $A_0$ are obtained as
\bea
A_1&=&2C_F\biggl[1-\dfrac{k_\beta}{2\sqrt{\Delta}}\ln\dfrac{k_\b+\sqrt{\Delta}}{k_\b-\sqrt{\Delta}}\biggr], \\
A_0&=&2C_F\bigg[-2\ln\d_s +\dfrac{1}{2\b_c}\ln\dfrac{1+\b_c}{1-\b_c} +\dfrac{1}{2\b_{\bar{c}}}\ln\dfrac{1+\b_{\bar{c}}}{1-\b_{\bar{c}}} \NO\\
&&+ \dfrac{k_\beta}{\sqrt{\Delta}}\ln\d_s\ln\dfrac{k_\b+\sqrt{\Delta}}{k_\b-\sqrt{\Delta}} -I(\b_c,\b_{\bar{c}},\cos\theta_{c\bar{c}})\biggr],\NO
\eea
where $\beta_{c(\bar{c})}$ is the ratio of momentum to energy of $c(\bar{c})$ and $\theta_{c\bar{c}}$ is the
angel between the $c\bar{c}$ pair. $k_\beta$ and $\Delta$ are defined as
\bea
k_\beta&=&1-\b_c\b_{\bar{c}}\cos\theta_{c\bar{c}}, \NO\\
\Delta&=&k_\b^2-(1-\b_c^2)(1-\b_{\bar{c}}^2)
\eea
and
\be
I(\b_c,\b_{\bar{c}},\cos\theta)=\int\limits_0^1 dx\dfrac{1}{f(x)[1-f(x)^2]}\ln\dfrac{1+f(x)}{1-f(x)},
\ee
with 
\be
f(x)=\biggl[(1-x)^2\b_c^2+2x(1-x)\b_c\b_{\bar{c}}\cos\theta+x^2\b_{\bar{c}}^2\biggr]^{1/2}.
\ee
The other way is to do the calculation in the CM frame of the open $c\overline{c}$ pair, and it leads to much simpler expressions for $A_1$ and $A_0$ as 
\bea
A_1&=&2C_F\biggl[1-\dfrac{1+\b^2}{2\b}\ln\dfrac{1+\b}{1-\b}\biggr], \NO\\
A_0&=&2C_F\bigg[-2\ln\d_s+\dfrac{1}{\b}\ln\dfrac{1+\b}{1-\b} -\dfrac{1+\b^2}{\b}\\
&&\times\biggl(\li\dfrac{2\b}{1+\b}+\dfrac{1}{4}\ln^2\dfrac{1+\b}{1-\b}-\ln\d_s\ln\dfrac{1+\b}{1-\b}\biggr)\biggr],\NO
\label{eqn:accbar}
\eea
where 
\be \b=\dfrac{|p_4|}{E_4}=\dfrac{|p_5|}{E_5}=\sqrt{1-\dfrac{1}{\hatscc}}=\dfrac{\Delta_1}{\Delta_2}\ee
is the ratio of momentum to energy for $c$ or $\overline{c}$ in the CM frame of $c\overline{c}$. The hard noncollinear part $\s^{H\overline{C}}$ is IR finite and can be numerically computed using standard Monte-Carlo integration techniques. The expressions for $A_{0,1}$ in the CM frame of $c\bar{c}$ in Eq.~(\ref{eqn:accbar}) can also be found in Ref.~\cite{Harris:2001sx}. The real cross section $\s^R$ is frame independent and should be the same no matter in the CM frame of $e^+e^-$ or $c\bar{c}$. It is obviously checked in our numerical calculations. 

Finally, all the IR singularities are canceled analytically. After adding all the contribution together, the cross section at NLO is expressed as
\be
\s^{(1)}=\s^{(0)}\left\{1+\dfrac{\a_s(\mu)}{\pi}\left[a(\hats)+\b_0\ln\left(\dfrac{\mu}{2m_c}\right)\right]\right\},
\label{eqn:CS_NLO} \ee where $\b_0$ is the one-loop coefficient of
the QCD beta function.
\begin{table}[htbp]
\begin{center}
\begin{tabular}{|c|c|c|c|c|c|c|}
\hline\hline
$m_c$(GeV)&$\a_s(\mu)$&$\s^{(0)}$(pb)&$a(\hats)$&$\s^{(1)}$(pb)&$\s^{(1)}/\s^{(0)}$\\
\hline
1.4&0.267&0.224&8.19&0.380&1.70 \\
\hline
1.5&0.259&0.171&8.94&0.298&1.74\\
\hline
1.6&0.252&0.129&9.74&0.230&1.78\\
\hline\hline
\end{tabular}
\caption{Cross sections with different charm quark mass $m_c$ with the renormalization scale 
$\mu=2m_c$ and $\sqrt{s}=10.6 \gev$.  $a({\hat s})$ is defined in Eq.~(\ref{eqn:CS_NLO}). }
\label{table:result}
\end{center}
\end{table}

To study the polarization of $\jpsi$ production, we define
the angular distribution coefficient $A$ as the $\a$ in Eq. (2.1) of Ref.~\cite{Cho:1996cg}:
\be
\dfrac{d^2\s}{d\cos\theta dP_\jpsi}=S(P_\jpsi)[1+A(P_\jpsi)\cos \theta],
\label{eqn:def:a}
\ee
and the polarization factor $\alpha$ is defined as
\be
\alpha(P_\jpsi)=\dfrac{d\sigma_T/dP_\jpsi-2d\sigma_L/dP_\jpsi} {d\sigma_T/dP_\jpsi-2d\sigma_L/dP_\jpsi},
\ee
where $P_\jpsi$ and $\theta$ are the 3-momentum and production angle of $\jpsi$ in the laboratory frame.
$\sigma_T$ and $\sigma_L$ 
are the transverse and longitudinal polarized cross section.
To calculate $\alpha(P_\jpsi)$, we use the same method to represent the polarized cross section as Eqs.~(8) and
(9) in Ref.~\cite{Gong:2008hk}. This method is found numerically
unstable in a small region of phase space due to the cancellation of
large numbers. Therefore, the momentum distributions for $A$ and
$\alpha$ contain potentially large numerical errors in our calculation for $P_\jpsi<1$ GeV or
$P_\jpsi>4.2$ GeV. As regards the total cross
section and momentum distribution of $J/\psi$ production, a
simplified method (see more details in Appendix.~\ref{chapter:guv}) is used to calculate the amplitude square with
very good convergence behavior in numerical calculations. But it cannot be applied to the calculation of $A$, $\alpha$ and $\cos\theta$ distributions.

In the NLO calculation, we should adopt $\a_s$ in the two-loop formula 
\be
\dfrac{\a_s(\mu)}{4\pi} =\dfrac{1}{\b_0 \ln(\mu^2/\Lambda_{QCD}^2)} - \dfrac{\b_1 \ln \ln(\mu^2/\Lambda_{QCD}^2)}{\b_0^3 \ln^2(\mu^2/\Lambda_{QCD}^2)},
\ee
with number of active quark flavors $n_f=4$ and $\Lambda^{(4)}_{\overline{\mathrm{MS}}}=0.338 \gev$.
The value of the wave function at the origin of $\jpsi$ is extracted from the leptonic decay widths:
\be
\Gamma_{ee}=\left(1-\dfrac{16}{3}\dfrac{\a_s}{\pi}\right) \dfrac{4\alpha^2e_c^2}{M_{\jpsi}^2}|R_s^\jpsi(0)|^2 .
\label{eqn:R0}
\ee
\begin{figure}
\center{
\includegraphics*[scale=0.4]{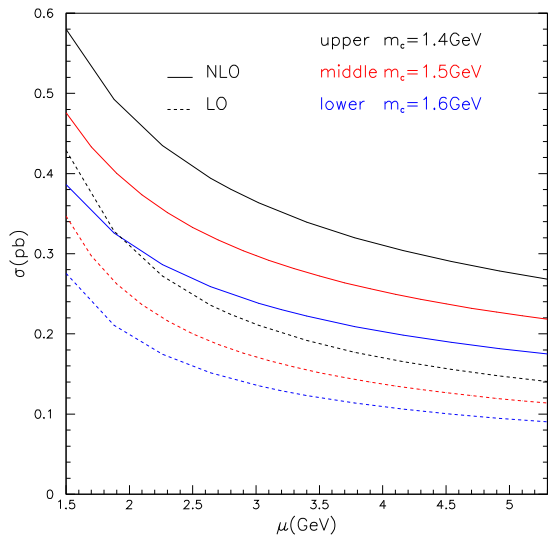}
\caption {\label{fig:scale1}Cross sections of $e^+e^-\rightarrow \jpsi c\bar{c}$ as a function of the renormalization scale $\mu$. The mass of charm quark is chosen as 1.4, 1.5 and 1.6 GeV, respectively.}}
\end{figure}
\begin{figure}
\center{
\includegraphics*[scale=0.4]{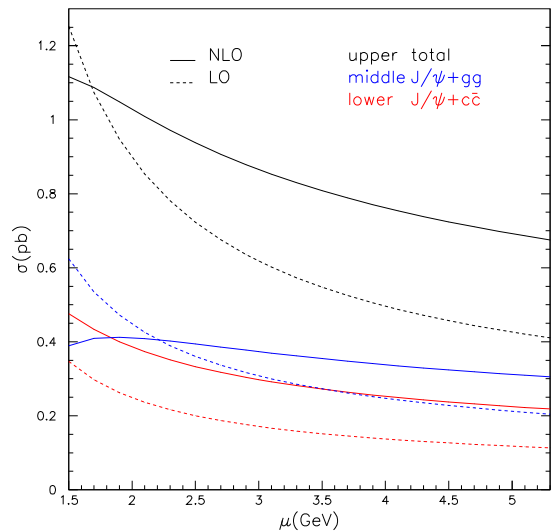}
\caption {\label{fig:scale2}Cross sections for $\jpsi c\bar{c}$, $\jpsi gg$ and total, as a function of the renormalization scale $\mu$.}}
\end{figure}
\begin{figure}
\center{
\includegraphics*[scale=0.4]{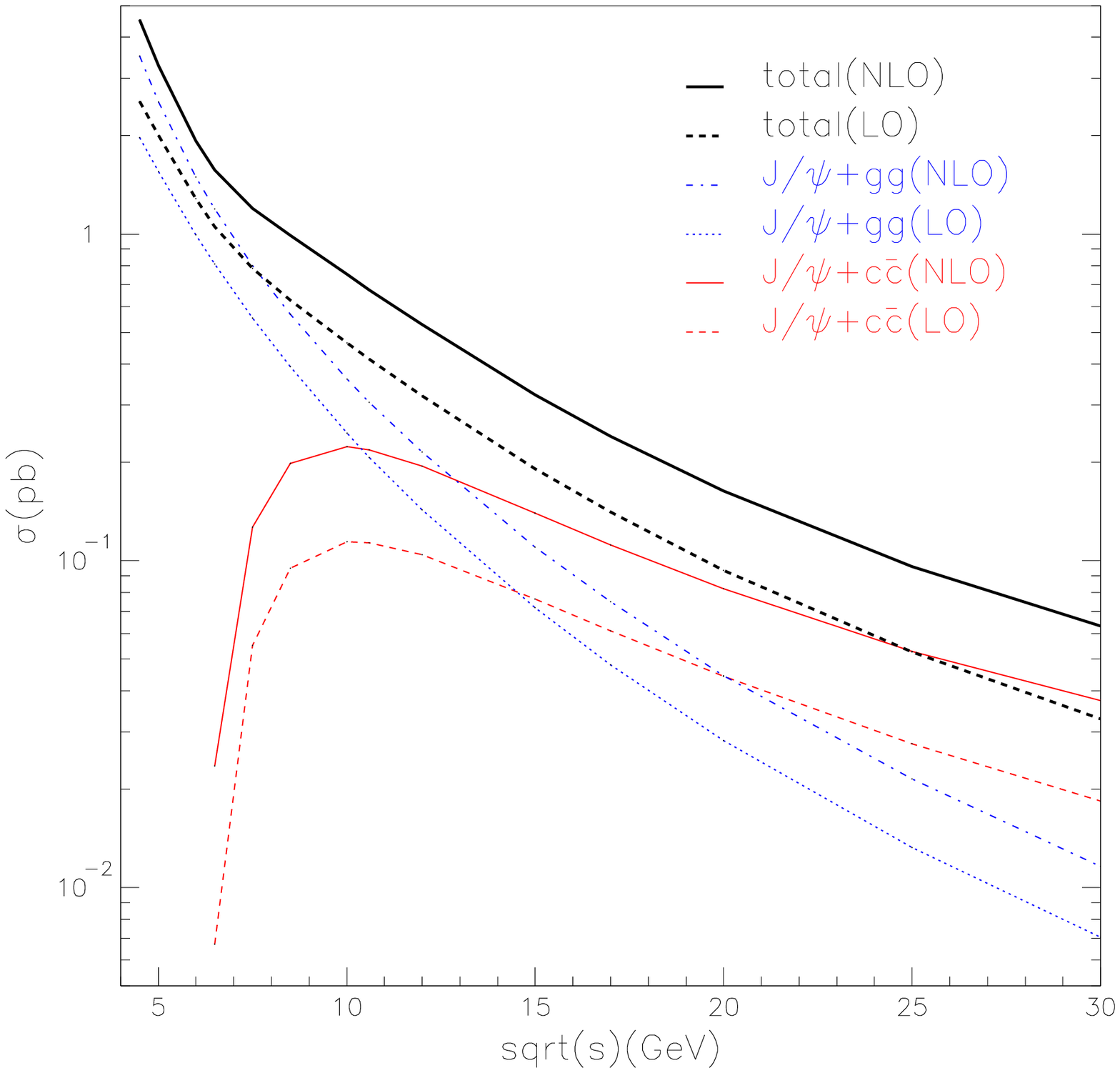}
\caption {\label{fig:sqrt_s}Cross sections for $\jpsi c\bar{c}$, $\jpsi gg$ and total, as a function of the center-of-mass energy of $e^+e^-$ $\sqrt{s}$ with the renormalization scale $\mu=\sqrt{s}/2$.}}
\end{figure}

\begin{figure}
\center{
\includegraphics*[scale=0.4]{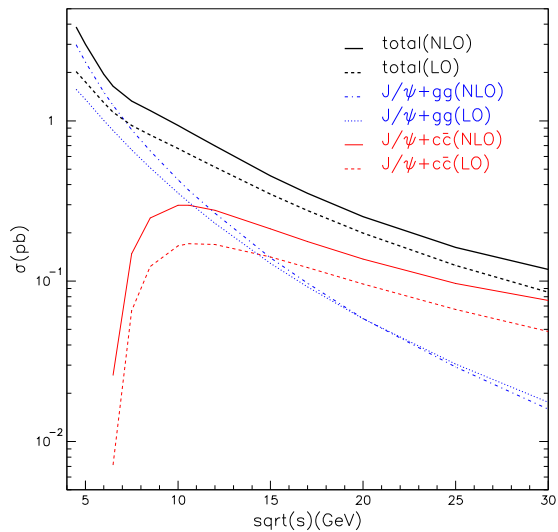}
\caption {\label{fig:sqrt_m}Cross sections for $\jpsi c\bar{c}$, $\jpsi gg$ and total, as a function of the center-of-mass energy of $e^+e^-$ $\sqrt{s}$ with the renormalization scale $\mu=2m_c$.}}
\end{figure}

\section{Numerical Results}
By using $\Gamma_{ee}=5.55\kev$, together with $\a=1/137$, $M_{\jpsi}=2m_c=3.0 \gev$ and $\a_s=0.26$, $|R_s(0)|^2=0.944 \gev^3$ is obtained. For other value of $m_c$, it should be multiplied by $(m_c/1.5 \gev)^2$. The numerical results are shown in Table.~\ref{table:result}. They are a bit smaller than those given by Zhang and Chao in Ref.~\cite{Zhang:2006ay}. It's just because of the differences in the value of $R_s(0)$ and renormalization scheme. If we choose their value of $R_s(0)$ and renormalization scheme, both calculations reach a very good agreement. Thus our calculation confirms the results in Ref.~\cite{Zhang:2006ay}. Thereafter, if not specified, we use $\sqrt{s}=10.6\gev$, $m_c=1.5\gev$ and $\mu=2m_c$ as our default choices in our results presented below. All of the results for $\jpsi gg$ are from our previous work~\cite{Gong:2009kp}. 

The scale dependence of the cross section is shown in Figs.~\ref{fig:scale1} and \ref{fig:scale2}. 
We can see from Figs.~\ref{fig:scale1} that the scale dependence of the total cross for 
$\jpsi c\bar{c}$ has not improved at NLO. 
The curves marked with "total" in Fig.~\ref{fig:scale2} denotes the combination of $\jpsi c\bar{c}$ 
and $\jpsi gg$ channels, together with the contribution from the feed-down of $\psi^\prime$ by 
multiplying a factor of 1.29. The treatment is applied to all the "total" results throughout 
this paper.
Figs.~\ref{fig:sqrt_s} and \ref{fig:sqrt_m} show the $\sqrt{s}$ dependence of the cross section, with $\mu=\sqrt{s}/2$ and $\mu=2m_c$, respectively. We see that the cross section of the $\jpsi c\bar{c}$ changes much milder than that of the $\jpsi gg$ channel as the center-of-mass energy increases.
The asymptotic behavior of the LO total cross section in the threshold region for both channels can be obtained easily as 
\bea
\s^{(0)}_{c\bar{c}}&=&\dfrac{8\a^2\a_s^2e_c^2|R_s(0)|^2}{27m_c^5}\times\dfrac{59\pi}{1024\sqrt{2}}\xi_{c\bar{c}}^3
+\co(\xi_{c\bar{c}}^4) \NO\\
\s^{(0)}_{gg} &=&\dfrac{\a^2\a_s^2e_c^2|R_s(0)|^2}{9m_c^5}\times\dfrac{8}{3}\xi_{gg} +\co(\xi_{gg}^2)
\eea
with $\xi_{c\bar{c}}=\sqrt{\hats}-2$ and $\xi_{gg}=\sqrt{\hats}-1$.
It is clearly shown that the production of $\jpsi c\bar{c}$ is through the p-wave channel and that of 
$\jpsi gg$ is through the s-wave channel near the threshold region. A simple $J^{PC}$ conservation analysis 
can easily explain these behavior. Even more, the threshold of the $\jpsi c\bar{c}$ channel is at 
$\sqrt{s}=4m_c$ while that of the $\jpsi gg$ channel is at $\sqrt{s}=2m_c$. 

\begin{figure}
\center{
\includegraphics*[scale=0.4]{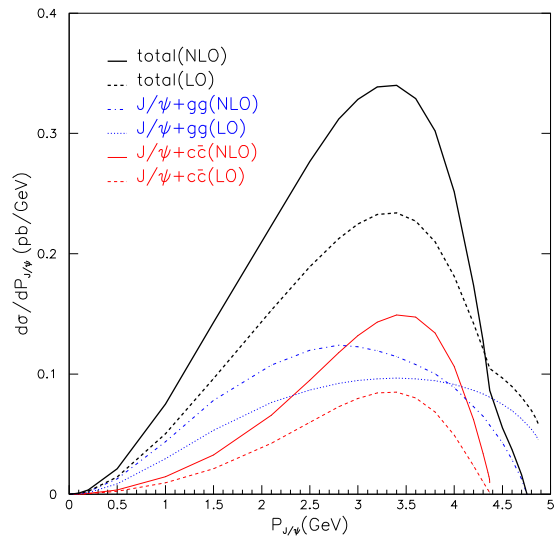}
\caption {\label{fig:p}Momentum distribution of $\jpsi$ production. The comparison with experimental
data will be made later.}}
\end{figure}

\begin{figure}
\center{
\includegraphics*[scale=0.4]{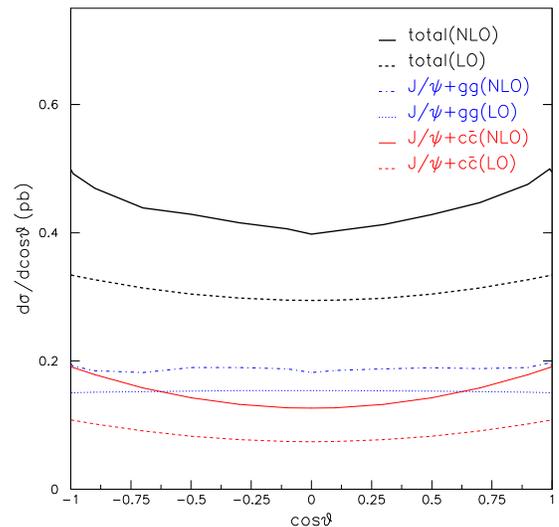}
\caption {\label{fig:cos}Angular distribution of $\jpsi$ production.
The comparison with experimental data will be made later.}}
\end{figure}

\begin{figure}
\center{
\includegraphics*[scale=0.4]{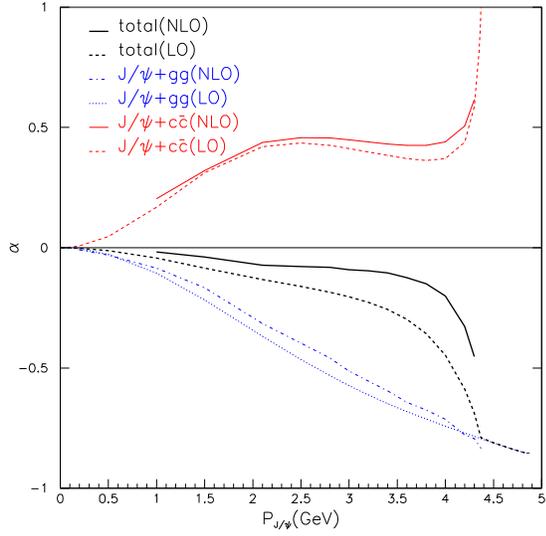}
\caption {\label{fig:polar}Momentum distribution of the polarization parameter $\alpha$ of $\jpsi$.}}
\end{figure}

\begin{figure}
\center{
\includegraphics*[scale=0.4]{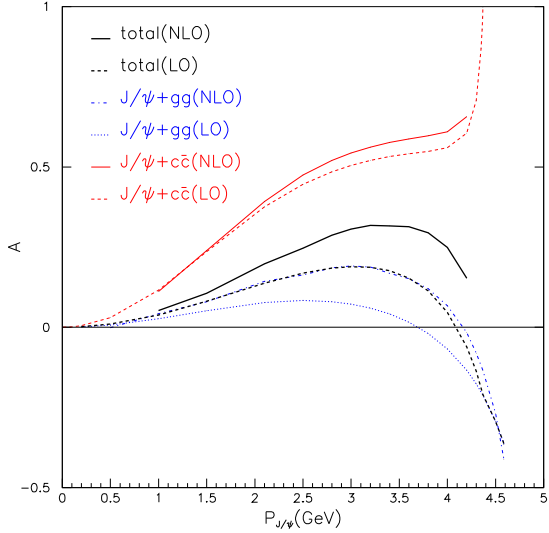}
\caption {\label{fig:a}Momentum distribution of the angular distribution parameter $A$ of $\jpsi$.}}
\end{figure}

The momentum and angular distributions of $\jpsi$ production are shown in
Figs.~\ref{fig:p} and \ref{fig:cos}.
It is found that the endpoint behavior of momentum distribution for the $\jpsi+gg$ channel was obviously
changed from LO to NLO due to a logarithm divergent term appearing in the NLO calculation. 
 We find that the shape of momentum distributions are similar with the recent experimental data~\cite{pakhlov:2009nj}, but the angular distributions are very different. Neither the $\jpsi c\bar{c}$ nor the $\jpsi gg$ channel can fit the experimental data.
In Figs.~\ref{fig:polar} and \ref{fig:a}, the momentum distributions of the
polarization factor $\a$ and the angular distribution coefficient
$A$ of $\jpsi$ are shown, and it can be seen that there is only a slight change for both $\alpha$ and $A$ from LO to NLO.  

\begin{figure}
\center{
\includegraphics*[scale=0.4]{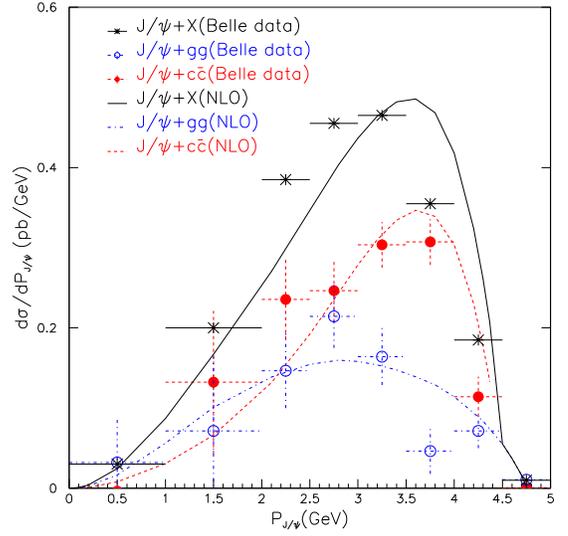}
\caption {\label{fig:p_new}Momentum distribution of inclusive $\jpsi$ production with $\mu=\mu^*$ and $m_c=1.4\gev$ is taken for the $\jpsi cc$ channel. The contribution from the feed-down of $\psi^\prime$ has been added to all curves by multiplying a factor of 1.29.}}
\end{figure}

\begin{figure}
\center{
\includegraphics*[scale=0.4]{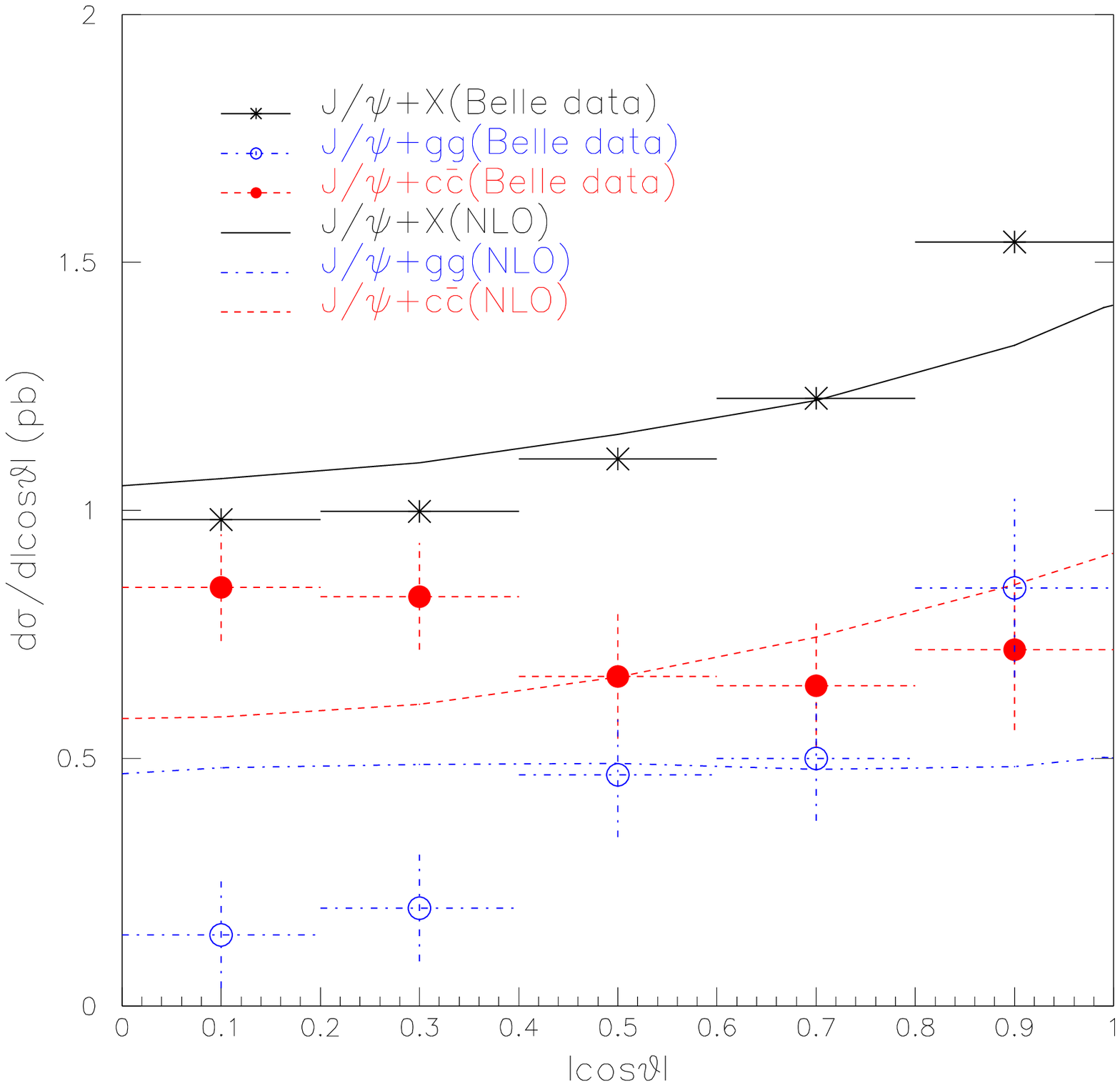}
\caption {\label{fig:cos_new}Angular distribution of inclusive $\jpsi$ production with $\mu=\mu^*$ and $m_c=1.4\gev$ is taken for the $\jpsi cc$ channel. The contribution from the feed-down of $\psi^\prime$ has been added to all curves by multiplying a factor of 1.29.}}
\end{figure}
\begin{figure}
\center{
\includegraphics*[scale=0.23]{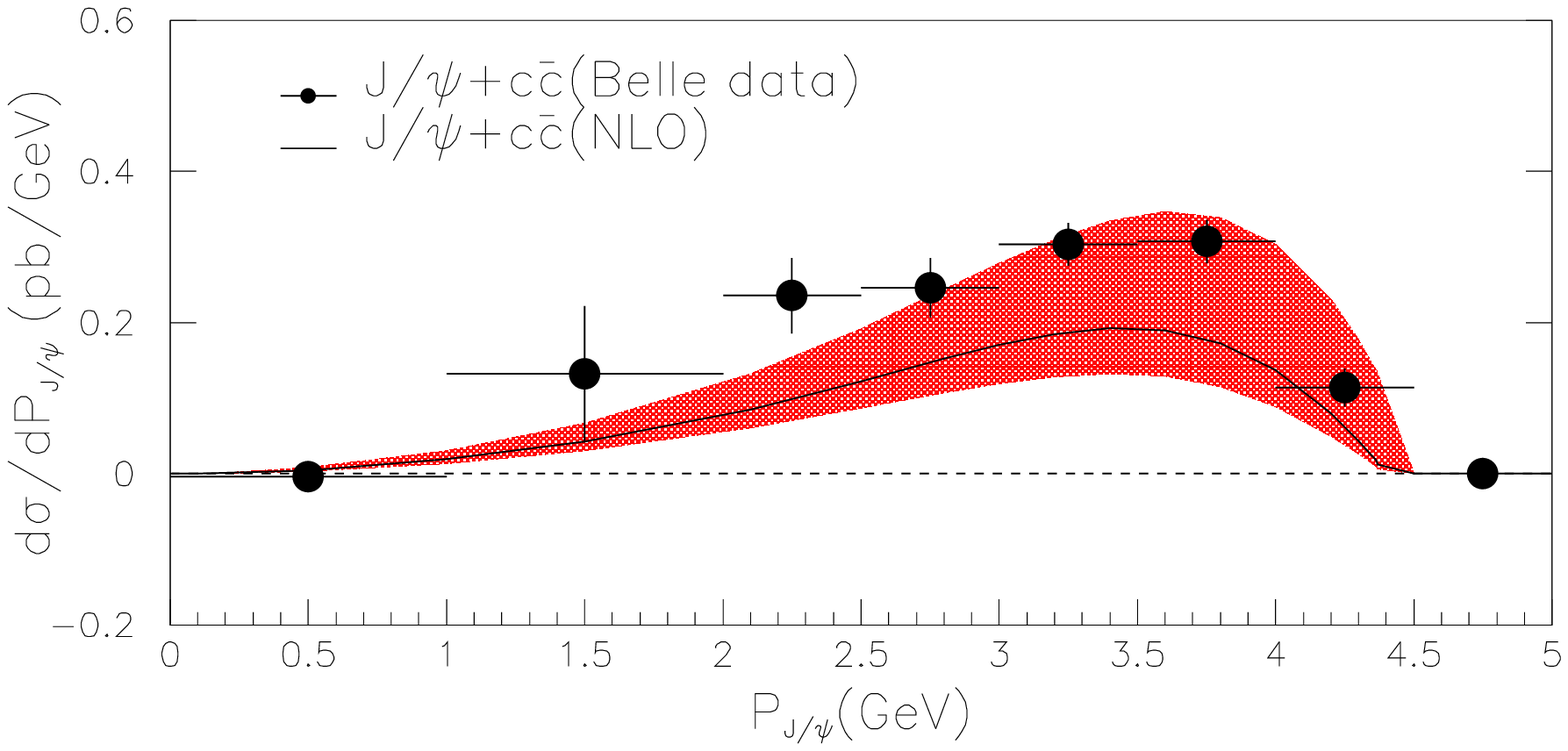}
\includegraphics*[scale=0.23]{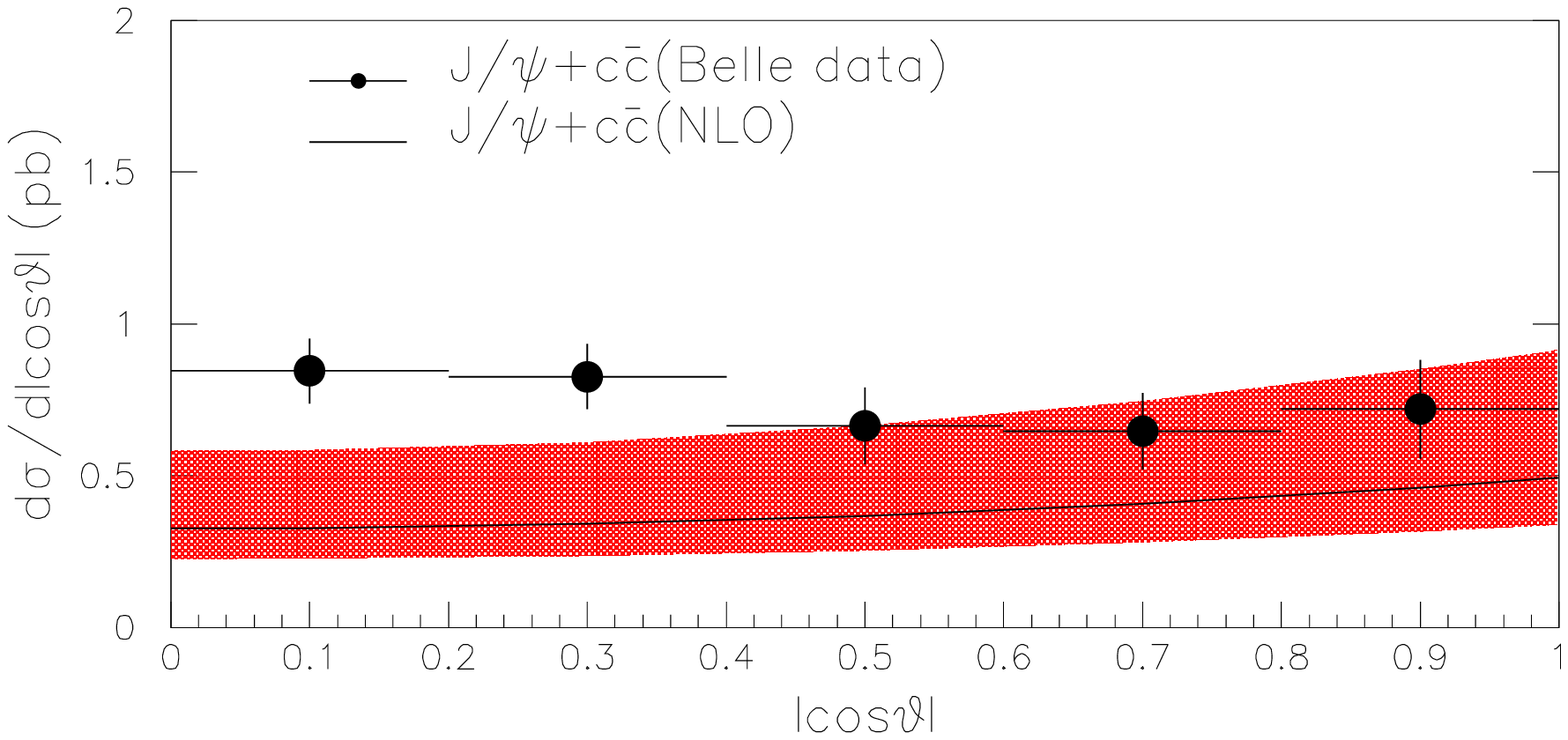}\\
\includegraphics*[scale=0.23]{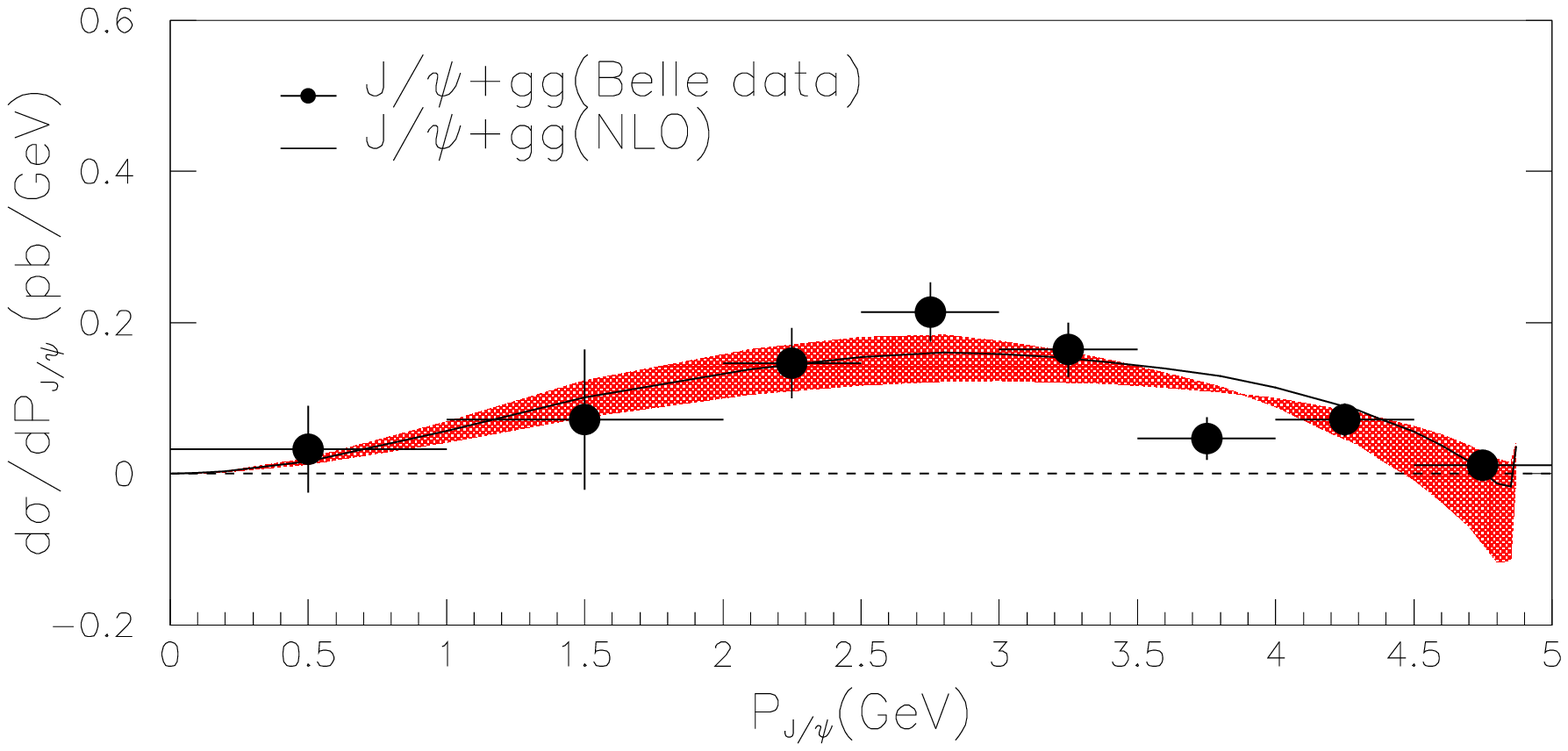}
\includegraphics*[scale=0.23]{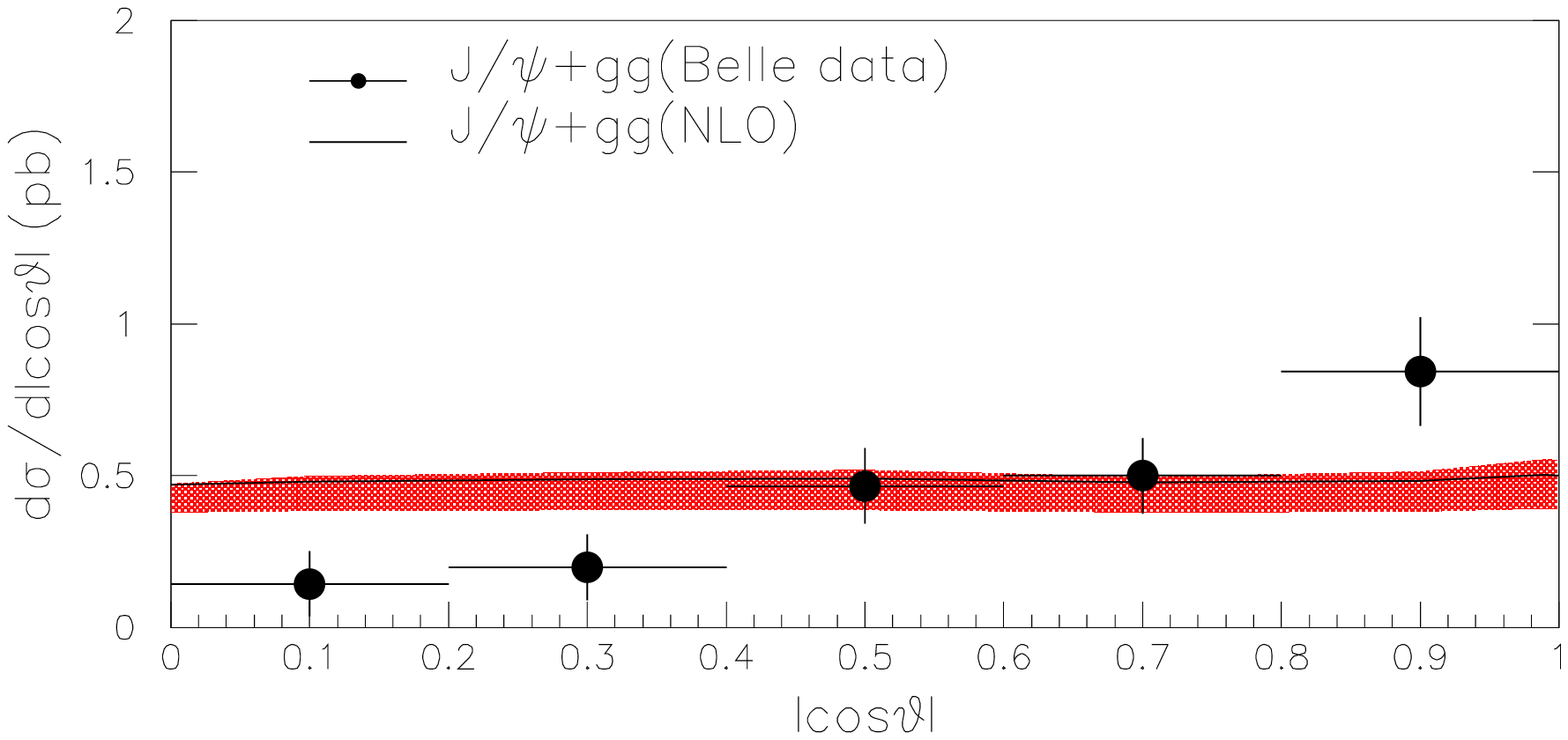}\\
\includegraphics*[scale=0.23]{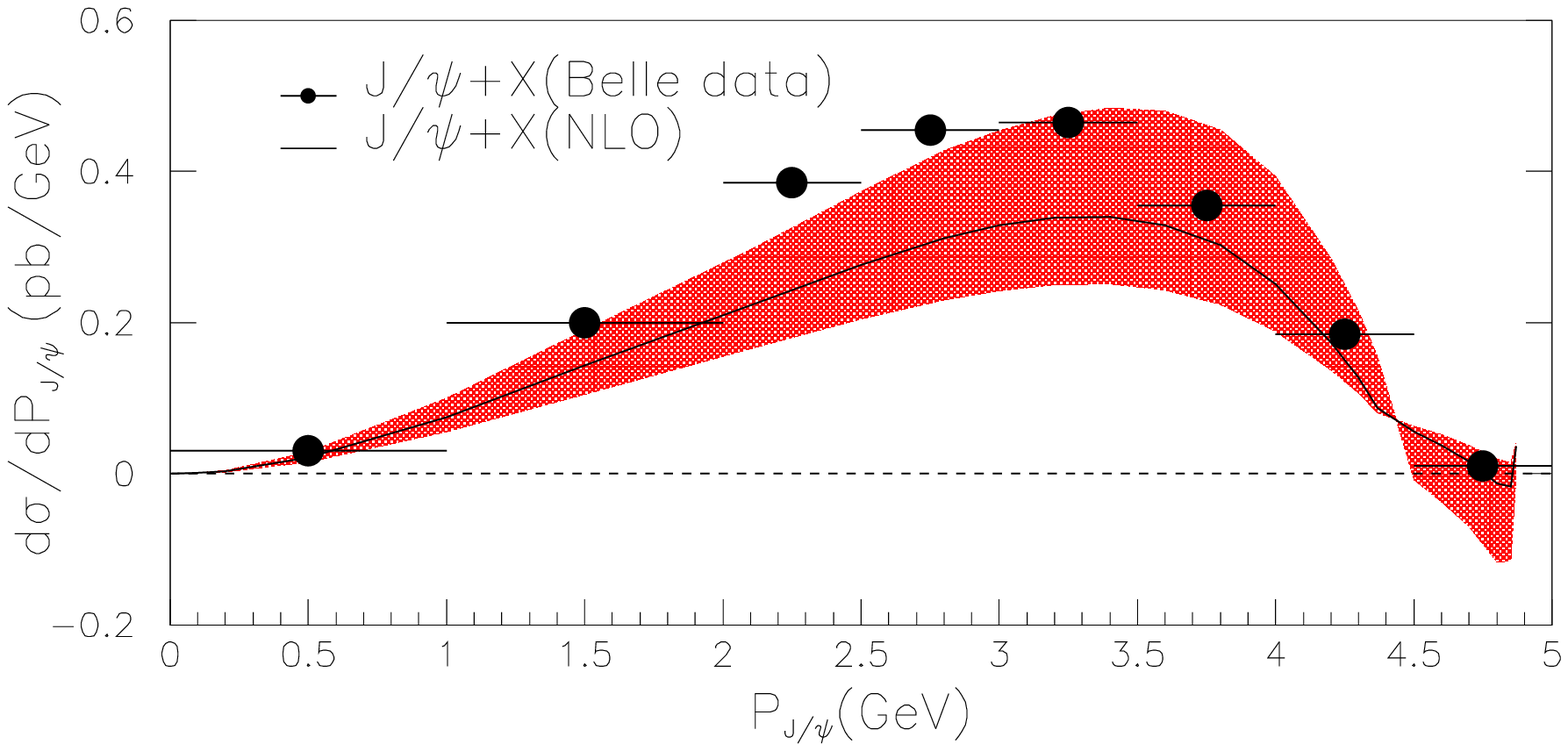}
\includegraphics*[scale=0.23]{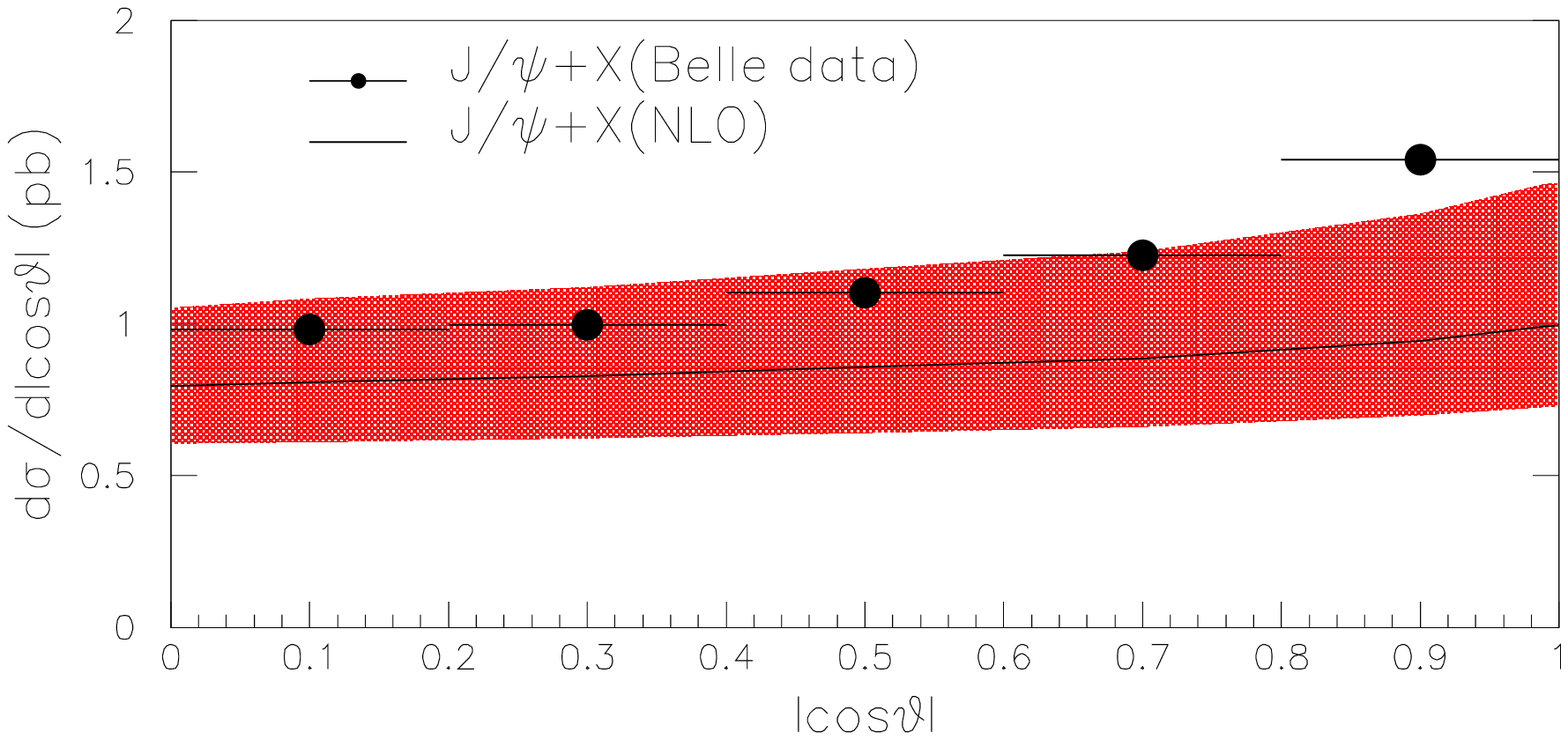}
\caption {\label{fig:band}Momentum and angular distributions of inclusive $\jpsi$ production. The contribution from the feed-down of $\psi^\prime$ has been added to all curves by multiplying a factor of 1.29. }}
\end{figure}

\section{More on the renormalization scale choice}
One possible way, although debatable,
to choose the renormalization scale is by following the procedures of Brodsky, Lepage and Mackenzie (BLM) scale setting~\cite{Brodsky:1982gc}; a unique scale choice $\mu^*$ is obtained and the cross section at NLO is expressed as
\be
\s^{(1)}=\s^{(0)}(\mu^*)[1+\frac{\a_s(\mu^*)}{\pi} b(\hat s)].
\label{eqn:BLM}
\ee
\begin{table}[htbp]
\begin{center}
\begin{tabular}{|c|c|c|c|c|c|c|c|}
\hline\hline
$m_c$(GeV)&$\a_s(\mu^*)$&$\s^{(0)}$(pb)&$b(\hats)$&$\s^{(1)}$(pb)&$\s^{(1)}/\s^{(0)}$&$\mu^*$(GeV)\\
\hline
1.4&0.348&0.381&3.77&0.540&1.42&1.65 \\
\hline
1.5&0.339&0.293&4.31&0.429&1.47&1.72\\
\hline
1.6&0.332&0.222&4.90&0.337&1.52&1.79\\
\hline\hline
\end{tabular}
\caption{Cross sections with different charm quark mass $m_c$. The renormalization scale 
$\mu=\mu^*$ is obtained by using BLM scale setting \cite{Brodsky:1982gc}, 
and $b({\hat s})$ is defined in Eq.~(\ref{eqn:BLM})}
\label{table:result2}
\end{center}
\end{table}

From the relevant results listed in Table.~\ref{table:result2},
we can see that the K factors become smaller and the convergence for QCD perturbative expansions 
is improved. In the case of $m_c=1.4$ GeV, the total cross section is 0.381 pb at LO and 
0.540 pb at NLO with the K factor 1.42.  To further consider the contribution from 
$\psi^\prime$ feed-down by introducing a factor $1.29$, the cross section is 0.70 pb at NLO, 
To include the contribution from $e^+e^-\rightarrow 2\gamma^*\rightarrow J/\psi c\bar{c}$ given in 
Ref~\cite{Liu:2003zr}, we should add about 0.03 pb and the total cross section is 0.73 pb now. 
The result can well explain the recent experimental measurement $0.74\pm0.08^{+0.09}_{-0.08}$ pb 
given by the Belle collaboration~\cite{pakhlov:2009nj}. It should be mentioned that the optimal scale 
choice $\mu^*=1.65$ GeV is close to $M_\jpsi/2$, half of the typical hard scale in the process.

In Figs.~\ref{fig:p_new} and \ref{fig:cos_new}, the momentum and angular distributions of inclusive $\jpsi$ are shown again, while this time $\mu=\mu^*$ and $m_c=1.4$ GeV is taken for the $\jpsi c\bar{c}$ channel. It is clearly shown in Fig.~\ref{fig:p_new} that the momentum distribution roughly fits the experimental data and for the $\jpsi gg$ part, the measurements at $P_\jpsi =2.75$ and $3.75$GeV
are not consistent with the theoretical calculations within the experimental error.  
For angular distributions shown in Fig.~\ref{fig:cos_new}, we see that the predication for the total 
angular distribution agrees rather well with experimental measurement,
but neither the $\jpsi c\bar{c}$ nor the $\jpsi~\mathrm{non}(c\bar{c})$ channel can fit the experimental measurements. In Fig.~\ref{fig:band}, a band of each line given in Figs.~\ref{fig:p_new} and \ref{fig:cos_new} is shown. The bands are obtained by verifying the renormalization scale and charm quark mass used in the calculation. For the $\jpsi cc$ channel, the boundaries of bands are given by the choices $\mu=\mu^*$, $m_c=1.4\gev$ and $\mu=4m_c$, $m_c=1.5\gev$. For the $\jpsi gg$ part, they are determined by $m_c\leq \mu \leq 4m_c$ with $m_c=1.5\gev$. We see that the situation for the two channels is improved with these theoretical uncertainty bands when compared 
with experimental measurements.

\section{Summary and Discussions}
We calculate the next-to-leading-order (NLO) QCD correction to $e^+e^-\rightarrow \jpsi c \bar{c}$ at the B factories, and present theoretical predictions on the momentum and production angular distribution for $J/\psi$ production, and momentum distribution for $J/\psi$ polarization at the NLO for the first time. It increases the cross section from 0.171 pb to 0.298 pb with a K factor of about $1.74$ for the default choice $m_c=1.5\gev$ and $\mu=2m_c$. By considering its dependence on the charm quark mass and renormalization scale with $\mu=2m_c$, the NLO cross section ranges from $0.230$ to $0.380\pb$. Furthermore, it will be enhanced by another factor of about $1.29$ when the feed-down from $\psi^\prime$ is considered. The total cross section agrees with that given by Zhang and Chao~\cite{Zhang:2006ay} when their renormalization scheme and input parameters are chosen. To further discuss the renormalization scale dependence, we applied the BLM scale setting~\cite{Brodsky:1982gc} for the renormalization scale and find that it improves the QCD perturbative expansion with the unique scale choice $\mu^*=1.65$ GeV and the K factor changes from 1.70 to 1.42 for $m_c=1.4$ GeV. Together with $\psi^\prime$ feed-down contribution, the total cross section ($0.73$pb) and momentum distribution can account for the recent experimental measurement~\cite{pakhlov:2009nj} when $m_c=1.4$ GeV and $\mu^*=1.65$ GeV are used. The total cross section and momentum distribution for the $e^+e^-\rightarrow\jpsi gg$ channel are also found to be consistent with the experimental measurement in previous studies~\cite{Ma:2008gq,Gong:2009kp}.  However, the production angular distribution for $\jpsi$ production for either the $\jpsi c\bar{c}$ or the $\jpsi gg$ channel has a quite different shape in contrast with the new experimental data, although it agrees with the experimental data when these two channels are added together. This situation is difficult to explain. 

To cut down the background from the very large electromagnetic $\jpsi$ production~\cite{Chang:1997dw}, more than four charged tracks are required in Belle's measurement. The measured value for the $\jpsi gg$ channel 
is smaller than the real value and the correction to this effects is difficult to perform. This incomplete
measurement could introduce uncertainty not only for $\jpsi gg$ total cross section but also for angular 
distribution and momentum distribution.
Therefore, the systematic errors for $\jpsi gg$ measurements were underestimated, and 
it is still
impossible to conclude that the color-octet contributions from $e^+e^-\rightarrow \jpsi^{(8)}(^1S_0,^3P_J)+g$ ~\cite{Braaten:1995ez} are ruled out since its momentum distribution can be changed under the resummation method by introducing a shape function~\cite{Fleming:2003gt}. To improve the measurement, it is better to cut the 
$\jpsi$ momentum at the large end point to cut down the large electromagnetic $\jpsi$ production, then the measurement will be a complete one without needing any correction. The better way to confirm or rule out the
color-octet prediction is to perform the measurement in the way suggested in Ref.~\cite{Wang:2003fw}.

To clarify the puzzle of polarization in $J/\psi$ production at the hadron collider, a detailed study on $J/\psi$ polarization at $e^+e^-$ will be very helpful. Therefore, further experimental measurements are strongly expected to testify our predictions on the momentum distribution for $J/\psi$ polarization. 

We thank Y. J. Zhang for helpful discussions. 
This work was supported by the National Natural Science Foundation of China (No.~10775141) and Chinese Academy of Sciences under Project No. KJCX3-SYW-N2.
\appendix
\section{A trick to simplify the calculation}\label{chapter:guv}
\begin{figure}
\center{
\includegraphics*[scale=0.4]{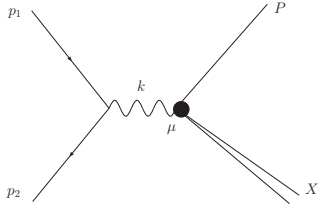}
\caption {\label{fig:method}Diagram of the process where the simplified method can be applied.}}
\end{figure}

Considering the process $e^+(p_1) +e^-(p_2)\rightarrow \gamma^*(k) \rightarrow H(P)+X$, where $X$ can be more than one particle, as show in Fig.~\ref{fig:method}. Thus the matrix element is expressed as $\mathrm{M}=l^\mu M_\mu$. Then the square of matrix element is expressed as $|\mathrm{M}|^2=L^{\mu\nu}M_{\mu}M_{\nu}^*$. Now define
\be M_{\mu\nu}^\prime\equiv\int[dP_X]M_{\mu}M_{\nu}^*, \ee 
where $[dP_X]$ denotes the integration over the momentums of all final state particles except H. Then $M_{\mu\nu}^\prime$ is expressed as
\be
M_{\mu\nu}^\prime=a_1P_\mu P_\nu +a_2 k_\mu k_\nu +a_3(P_\mu k_\nu +k_\mu P_\nu) +a_4 g_{\mu\nu}.
\ee
The current conversion demands
\be
k^\mu  M_{\mu\nu}^\prime = k^\nu  M_{\mu\nu}^\prime =0,
\ee
thus $M_{\mu\nu}^\prime$ is further expressed as
\bea
M_{\mu\nu}^\prime&=&c_1(k_\mu k_\nu - k^2 g_{\mu\nu}) \\
&&+c_2 [(k\cdot P)g_{\mu\alpha} - P_\mu k_\a][(k\cdot P) g_{\nu\alpha} - P_\nu k_\a].\NO
\eea
Also, we have
\be
L_{\mu\nu}=4(-\frac{1}{2}k^2g_{\mu\nu}-\frac{1}{2}q_\mu q_\nu +k_\mu q_\nu +q_\mu k_\nu),
\ee
where $k=p_1+p_2$ and $q=p_1-p_2$.
In the $p_1+p_2$ rest frame, the momentums are written as
\bea
k&=&p_1+p_2=(\sqrt{s},0,0,0), \NO\\
q&=&p_1-p_2=(0,0,0,\sqrt{s}), \NO\\
P&=&(E,0,P\sin\theta,P\cos\theta),
\eea
where $z$ axis is chosen along the beam direction. Then we have
\be
L^{\mu\nu} M_{\mu\nu}^\prime=2s^2[2c_1+c_2(P^2-2E^2-P^2\cos^2\theta)],
\label{eqn:luv}
\ee
from which one can obtain Eq.~(\ref{eqn:def:a}).
On the other hand, if we replace the $L^{\mu\nu}$ on the left side of Eq.~(\ref{eqn:luv}) with $-g^{\mu\nu}$, we have
\be 
-g^{\mu\nu} M_{\mu\nu}^\prime=s[3c_1+c_2(P^2-3E^2)].
\ee
Do the integration over $\theta$ and we have
\bea
\int d\cos\theta L^{\mu\nu} M_{\mu\nu}^\prime &=&8s^2[c_1+c_2(\frac{1}{3}P^2-E^2)] \propto \dfrac{d\s}{dE}\NO\\
\int d\cos\theta (-g^{\mu\nu})M_{\mu\nu}^\prime&=&6s[c_1+c_2(\frac{1}{3}P^2-E^2)]  
\label{eqn:luv_guv} \\
&=&\dfrac{3}{4s}\int d\cos\theta L^{\mu\nu} M_{\mu\nu}^\prime. \NO
\eea
Eq.~(\ref{eqn:luv_guv}) shows that, in the calculation of momentum distribution or total cross section, we can replace the $L^{\mu\nu}$ at the matrix elements squared level with $-4sg^{\mu\nu}/3$ to simplify the calculation.
\bibliography{paper}
\end{document}